\documentclass[aps,pre,nofootinbib,twocolumn,showpacs,superscriptaddress]{revtex4-1}
\usepackage[usenames,dvipsnames]{color,xcolor}
\usepackage{graphicx,epsfig,amssymb,amsmath,hyperref}
\usepackage[normalem]{ulem}
\hypersetup{colorlinks, citecolor=Red, linkcolor=Blue, urlcolor=Red}

\begin{document}
\title{Dynamic wettability alteration in immiscible two-phase flow in porous
  media: Effect on transport properties and critical slowing down}

\author{Vegard Flovik}

\email{vegard.flovik@ntnu.no}

\affiliation{Department of Physics, Norwegian University of Science
  and Technology, N-7491 Trondheim, Norway.}

\author{Santanu Sinha}

\email{santanu.sinha@ntnu.no}

\affiliation{Department of Physics, University of Oslo, P. O. Box 1048
  Blindern, N-0316 Oslo, Norway.}

\author{Alex Hansen}

\email{alex.hansen@ntnu.no}

\affiliation{Department of Physics, Norwegian University of Science
  and Technology, N-7491 Trondheim, Norway.}

\date{\today}

\begin{abstract}
The change in contact angles due to the injection of low salinity
water or any other wettability altering agent in an oil-rich porous
medium is modeled by a network model of disordered pores transporting
two immiscible fluids. We introduce a dynamic wettability altering
mechanism, where the time dependent wetting property of each pore is
determined by the cumulative flow of water through it. Simulations are
performed to reach steady-state for different possible alterations in
the wetting angle ($\theta$). We find that deviation from oil-wet
conditions re-mobilizes the stuck clusters and increases the oil
fractional flow. However, the rate of increase in the fractional flow
depends strongly on $\theta$ and as $\theta\to 90^\circ$, a critical
angle, the system shows critical slowing down which is characterized
by two dynamic critical exponents.
\end{abstract}


\maketitle

\section{Introduction}\label{introduction}
The world's primary energy demand is predicted to increase by
one-third between 2011 and 2035, where $82\%$ of it comes from fossil
fuels \cite{WEO13}. In this scenario, the fact that some $20$ to $60$
percent of the oil remains unrecovered in a reservoir after the
production is declared unprofitable, is a challenge of increasing
importance \cite{oilreserves}. The main reason for this loss is the
formation of oil clusters trapped in water and held in place by
capillary forces, which in turn are controlled by the wetting
properties of the reservoir fluids with respect to the matrix
rock. The production from complex oil reserves that today are
considered immobile or too slow compared to the cost is therefore an
important area of research. In this context, the role of formation
wettability is a focus area within the field of Enhanced Oil Recovery
(EOR) \cite{schlumberger}.

Different reservoir rocks have widely different wetting
characteristics \cite{sshv07}. Wettability may vary at the pore level
from strongly oil wet through intermediate wetting to strongly water
wet. Carbonate reservoirs contain more than half of the world's
conventional oil reserves, but the oil recovery factor is very low
compared to sandstone reservoirs \cite{sheng13}. This is due to the
complex structure, formation heterogeneity and more chemically active
wettability characteristics of the carbonate reservoirs, which leads
to uncertainty in the fluid flow and oil recovery
\cite{cy83}. Sandstone is strongly water wet before oil migrates from
a source rock into the reservoir. When oil enters a pore, it displaces
the water which leaves behind a water film sandwiched between the oil
and rock surface. This happens as a result of balancing van der Waals
and electric double layer forces, capillary pressure and grain
curvature \cite{thinfilm1}. A permanent wettability alteration is then
believed to take place by adsorption of asphaltenes from the crude oil
to the rock, and leads to high but slow recovery through continuous
oil films \cite{oilfilms1,oilfilms2}. As the oil saturation drops,
these films can become discontinuous, leaving immobile oil clusters
held in place by capillary forces.

Wettability is therefore an important petrophysical property which
plays a key role in the fluid transport properties of both
conventional and unconventional reservoirs \cite{ulr}, and there is
great potential to improve the oil recovery efficiency by altering the
wetting properties. Main factors which can alter the pore wettability
are; lowering the salinity \cite{EOR}, adding water-soluble
surfactants \cite{EOR2,mohan11} or adding oil-soluble organic acids or
bases \cite{WEOR}. Increasing the reservoir temperature also increases
water-wetness \cite{schembre06,sshv07}. There are some correlations
with the wetting behavior to the electrostatic forces between the rock
and oil surfaces \cite{elstatic}, but there is no consensus on the
dominating microscopic mechanism behind the wettability alteration. It
is known from laboratory experiments and field tests that a drift from
strongly oil-wet to water-wet or intermediate-wet conditions can
significantly improve the oil recovery efficiency \cite{WEOR}. The
amount of change in the wetting angle is a key factor here
\cite{wang01,wang02} which not only decides the increase in oil flow
but also the speed of the process. An improper change in the wetting
angle can also make the recovery very slow and not profitable.

Given there is a certain change in the wetting angle due to a brine,
the next important factor is the flow pathways in the matrix rock
which transports the oil and brine. One cannot expect any change in
the wetting angle of a pore if there is no flow of the brine through
it. The flow pathways depend on several different factors: the porous
network itself, oil saturation, capillary number and also on the
present wettability conditions. A change in the wettability will cause
a perturbation in the flow distribution of the system. This will in
turn again affect the wettability change through the altered flow
pathways, causing further changes in the flow distribution. The
dynamics of wettability alterations is therefore controlled by a
strongly correlated process.

There are some studies of wettability alterations in two-phase flow by
equilibrium-based network models \cite{b98} for capillary dominated
regimes where viscous forces are negligible. Moreover, investigation
of the time-scale of dynamics lacks attention in such models. In this
article, we present a detailed study of wettability alterations in
two-phase flow considering a network model of disordered pores
transporting two immiscible fluids where a dynamic wettability
alteration mechanism, correlated with the flow-pathways, is
implemented. We will focus on the transport properties due to the
change in the wettability as well as on the time scale of the
dynamics.

We study in the following the effect of wettability changes on
immiscible two-phase flow based on a network model
\cite{networkmodel,meniscimove,sinha11}. In section \ref{Model}, we
present the model and how we adapt it to incorporate the dynamic
wettability changes.  In Section \ref{Results}, we present our
results. Initially, we let the two phases settle into a steady state
where the averages of the macroscopic flow parameters no longer
evolve. At some point, we then introduce the wettability altering
agent, so that it starts changing the wetting angle. The wetting
angle alteration depends on the cumulative volume of the wettability
altering fluid that has flowed past a given pore. This induces
transient behavior in the macroscopic flow properties and we measure
the time it takes to settle back into a new steady state. We find
that there is a critical point at a wetting angle of 90 degrees and we
measure its dynamical critical exponents; the exponents are different
whether one approaches the critical point from smaller or larger
angles. In Section \ref{Conclusions} we summarize and conclude.

\section{Model}\label{Model}
\begin{figure}
\centerline{
\hfill\psfig{file=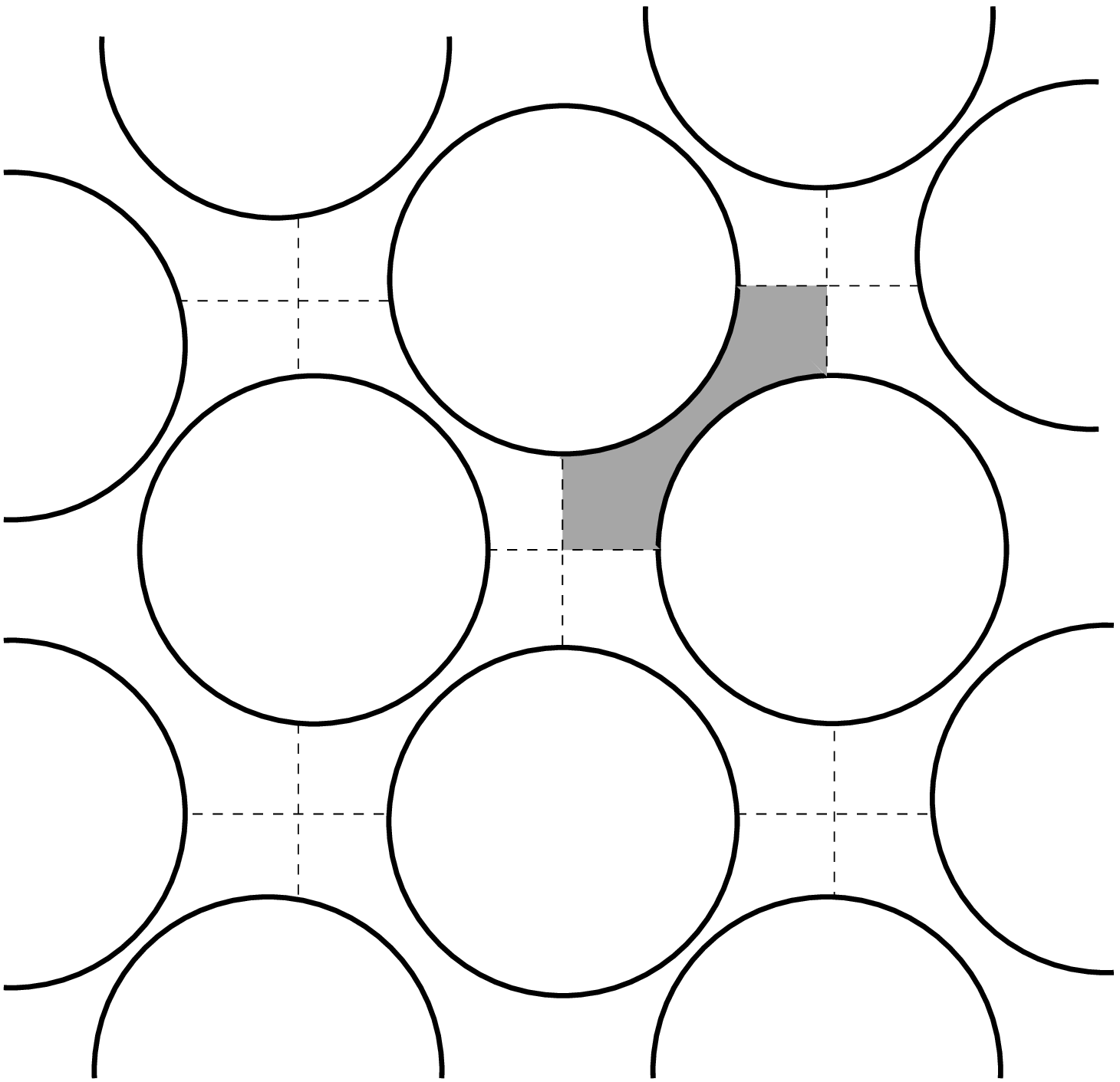,width=0.20\textwidth}\hfill
\hfill\psfig{file=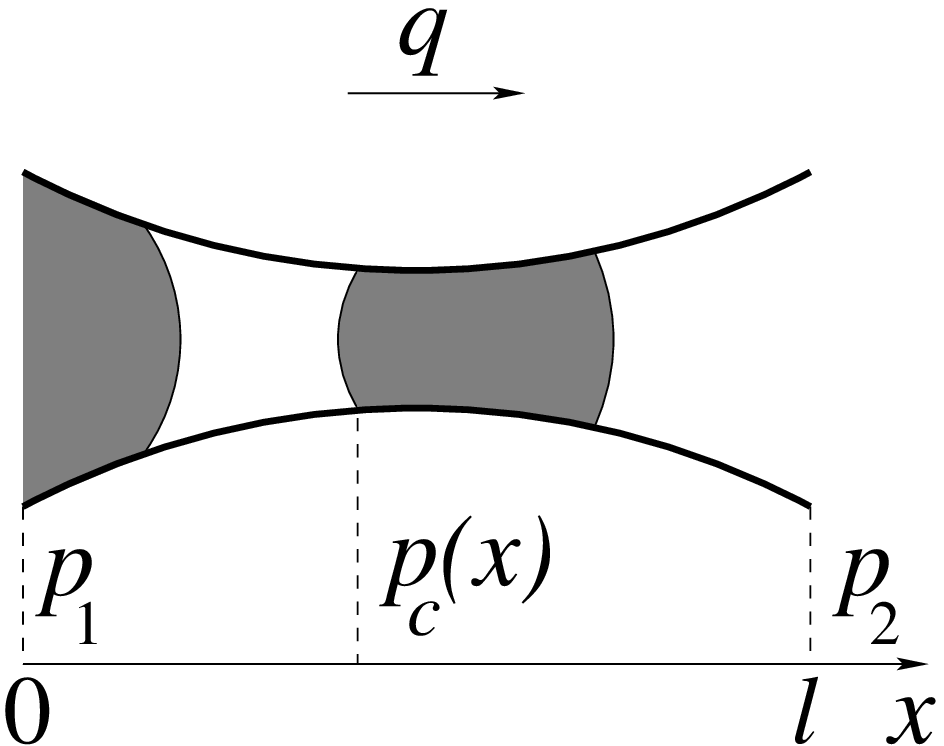,width=0.18\textwidth}\hfill}
\medskip
\centerline{\hfill (a) \hfill \hfill (b) \hfill}
\caption{\label{modelfig} (a) Illustration of the
  pore network model, constructed by links (tubes) with random radii
  connected to each other via nodes at the intersection of the dashed
  lines. One single link in the network is highlighted by gray which
  is again shown in (b) filled with two fluids. The wetting and
  non-wetting fluids are colored by white and gray respectively. $p_1$
  and $p_2$ are the pressures at the two ends of the link and $q$ is
  the local flow-rate. There are three menisci inside and capillary
  pressure difference at a menisci is indicated as $p_c(x)$.}
\end{figure}

We model the porous medium by a network of tubes (or links) oriented
at 45 degrees relative to the overall flow direction. The links
contain volumes contributed from both the pore and the throat, which
then intersect at volume-less vertices (or nodes). Any disorder can be
introduced in the model by a proper random number distribution for the
radius $r$ of each link, and we choose a uniform distribution in the
range $[0.1l,0.4l]$ here, where $l$ is the length of each tube. The
network transports two immiscible fluids (we name them as oil and
water), one of which is more wetting than the other with respect to
the pore surface. The pores are assumed to be in between particles,
and the pore shape is thus approximated to be hour-glass shaped,
which introduces capillary effects in the system. The model is
illustrated in figure \ref{modelfig}.

Due to the hour glass shape of the pore, the capillary pressure at a
menisci separating the two fluids is not constant, and depends on the
position $x$ of the menisci inside the pore. The capillary pressure
$p_c(x)$ at position $x$ inside the $i$th pore is then calculated from
a modified form of the Young Laplace equation
\cite{Dullien,networkmodel},
\begin{equation}
\displaystyle p_c(x) = \frac{2\gamma \cos\vartheta_i}{r_i} \left[1 -
  \cos\left(\frac{2\pi x}{l}\right)\right].
\label{ylapeq}
\end{equation}
where $\gamma$ is the interfacial tension between the fluids and
$\vartheta_i$ is the wetting angle for that pore. As an interface
moves in time, $p_c(x)$ changes. The capillary pressure is zero at the
two ends ($x=0$ and $l$) and it is maximum at the narrowest part of
the pore. It makes the model closer to the dynamics of drainage
dominated flow, where the film flow can be neglected. When there are
multiple menisci in a pore, the total capillary pressure inside the
$i$th pore is obtained from the vector sum ($\sum_i p_c(x)$) over all
the menisci in that pore. The flow is driven by maintaining a constant
total flow rate $Q$ throughout the network, which introduces a global
pressure drop. The instantaneous local flow rate $q_i$ inside the
$i$th link between two nodes with pressures $p_1$ and $p_2$ follows
the Washburn equation of capillary flow \cite{washburn},
\begin{equation}
\displaystyle
\label{wbeq}
q_i = -\frac{\pi r_i^2 k_i}{\mu_i^\text{eff}l} \left[\Delta p_i -\sum_i
p_c(x)\right],
\end{equation}
where $\Delta p_i=p_2-p_1$. $k_i=r_i^2/8$ is the permeability of
cylindrical tubes. Any other cross-sectional shape will only lead to
an additional overall geometrical
factor. $\mu_i^\text{eff}=\mu_\text{o}s_i+\mu_\text{w}(1-s_i)$, is the
volume averaged viscosity of the two phases inside the link, which is
a function of the oil saturation $s_i$ in that link. Here
$\mu_\text{o}$ and $\mu_\text{w}$ are the viscosities of oil and water
respectively.

The flow equations for the tube network are solved using a conjugate
gradient method \cite{fourieracc}. These are the Kirchhoff equations
balancing the flow, where the net fluid flux through every node should
be zero at each time step, combined with the constitutive equation
relating flux and pressure drop across each tube.  The system of
equations is then integrated in time using an explicit Euler scheme
with a discrete time step and all the menisci positions are changed
accordingly. Inside the $i$th tube, all menisci move with a speed
determined by $q_i$. Physical processes like bubble snap-off and
coalescence are introduced in the model. Due to these, bubbles can be
formed or merged, which changes the number of menisci inside a link
with time. When a meniscus reaches the end of a tube, new menisci are
formed in the neighboring tubes depending upon the flow rates. In each
link, a maximum number of menisci is allowed to form. When this number
is exceeded, the two nearest menisci are merged, keeping the volume of
each fluid conserved. In our simulations, we considered a maximum of
four menisci inside one pore which can be tuned depending on the
experimental observations. Further details regarding the menisci
movement can be found in Reference \cite{meniscimove}.

The simulations are started with an initial random distribution of two
fluids in a pure oil-wet network. Bi-periodic boundary conditions are
implemented in the system, which effectively makes flow on a torus
surface. The flow can therefore go on for infinite time, keeping the
saturation constant and the system eventually reaches to a steady
state. In the steady state, both drainage and imbibition take place
simultaneously and fluid clusters are created, merged and broken into
small clusters. One can consider this as the secondary recovery
stage. Once the system reaches the steady-state in a oil-wet network,
the dynamic wettability alterations are implemented, which may be
considered as the tertiary recovery stage or EOR. In the following we
discuss this in detail.

\subsection*{Dynamic wettability alteration}
\begin{figure}
\centerline{\hfill
\psfig{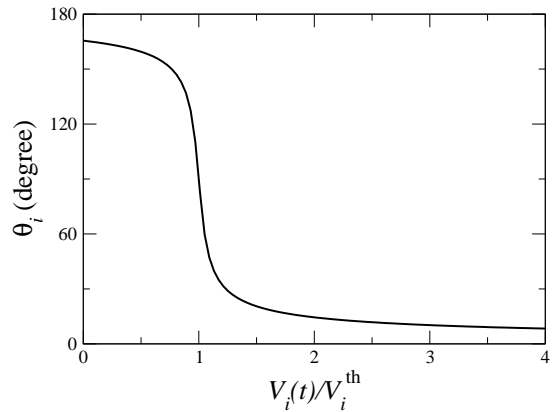}\hfill}
\caption{\label{distrfunc} Change in the wetting angle for a link
  given by $G_i(t)$ as a function of the cumulative volume of water
  $V_i(t)$ through that link.}
\end{figure}

We now introduce a dynamic wettability alteration mechanism to
simulate any wetting angle change, decided by the oil-brine-rock
combination and the distribution of the flow channels in the
system. In a previous study \cite{sinha11}, a simplified static
wettability alteration mechanism was studied, where the alteration
probability was considered equal for all pores without any correlation
with the flow of brine inside a pore. However, for wettability
alterations to occur, the wettability altering agent
(e.g. low-salinity water or surfactant) needs to be in contact with
the pore walls. Thus, the wettability alteration should follow the
fluid flow pathways and any change in the wetting angle inside a pore
should depend on the cumulative volumetric flow of brine through that
pore. This claim is rather trivial, as one can not expect any
wettability change in a pore if the altering agent is not
present. This means that if a certain pore is flooded by large amounts
of brine, the wetting angle should change more in that pore than the
one which had very little water flooded through. This is implemented
in the model by measuring the cumulative volumetric flux $V_i(t)$ in
each individual pore with time $t$,
\begin{equation}
\displaystyle
\label{flowsum}
V_i(t) = \sum_{\tilde{t}=t_0}^{t} q_i(\tilde{t})(1 -
s(\tilde{t}))\Delta \tilde{t},
\end{equation}
where $t_0$ is the time when the injection of low salinity water is
initiated, $\Delta \tilde{t}$ is the time interval between two
simulation steps and $(1-s(\tilde{t}))$ is the water
saturation. $V_i(t)$ is then used to change the wetting angle for each
tube continuously, updated at every time step after $t=t_0$. The
wetting angle $\vartheta_i$ of the $i$th pore can change continuously
from $180^\circ$ to $0^\circ$ as $V_i(t)$ changes from $0$ to
$\infty$. Correspondingly, the $\cos\vartheta_i$ term in
Eq. (\ref{ylapeq}) will change from $-1$ to $1$ continuously. This
continuous change of the wetting angle with the variation of $V_i(t)$
is modelled by a function $G_i(t)$ given by,
\begin{equation}
\displaystyle
\label{distr}
G_i(t) = \frac{2}{\pi}\tan^{-1}\left[C
  \left(\frac{V_i(t)}{V^\text{th}_i} - 1\right)\right]
\end{equation}
which replaces the $\cos\vartheta_i$ term in Eq. (\ref{ylapeq}). The
pre-factor $2/\pi$ is a normalization constant to set the range of the
function, and the parameter $C$ adjusts the slope during the
transition between oil wet and water wet. This leads to the change in
the wetting angle as a function of $V_i(t)$ as shown in
Fig. \ref{distrfunc}.

As our model does not include film flow, the wetting angle should not
reach either $0$ or $180$ degrees. Rather, a starting point of
$\vartheta_i \approx 165^\circ$ was used by choosing $C=20$ in our
simulations as shown in Fig. \ref{distrfunc}. Next, as a larger pore
will need more brine to be flooded in order to have a similar change
in the wetting angle than a smaller pore, a threshold value
$V^\text{th}_i$ is introduced, which is proportional to the volume of
that pore,
\begin{equation}
V_i^\text{th}=\eta\pi r_i^2l.
\label{vth}
\end{equation}
At $V_i(t)=V^\text{th}_i$, the wetting angle reaches to $90^\circ$ in
that pore and $p_c(x)$ essentially becomes zero. Here $\eta$ is a
proportionality constant which decides how many pore volumes of water
is needed to reach $V_i(t)=V^\text{th}_i$ for the $i$th pore. This
parameter can possibly be adjusted against future experimental
results, but is considered as a tuning parameter in this study. The
expression for the capillary pressure at a menisci from
Eq. \ref{ylapeq} then takes the form,
\begin{equation}
\displaystyle p_c(t) = \frac{2\gamma
  G_i(t)}{r_i}\left[1-\cos\left(\frac{2\pi x}{l}\right)\right].
\label{pcmod}
\end{equation}

The maximum amount of wetting angle that can be changed depends on the
combination of brine, oil and rock properties \cite{wang01,wang02}. We
therefore set a cut-off $\theta$ in the wetting angle change, such
that any pore that has reached to a wetting angle
$\vartheta_i=\theta$, can not be changed further. The model thus
includes all the essential ingredients of wettability alteration study
-- it is a time dependent model where the wettability alteration is
correlated with the flow pathways of the brine, and can be used to
study any oil-brine-rock combination decided by $\theta$.

\section{Results}\label{Results}

\begin{figure}
\centerline{\hfill
\psfig{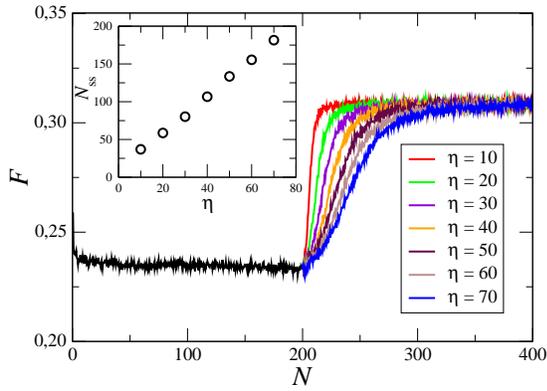}\hfill}
\caption{\label{eta} (Color online) Change in the oil-fractional-flow
  ($F$) during the simulation as a function of the fluid volume passed
  through any cross-section of the network expressed in terms of the
  number of pore-volumes ($N$). The network is of size $40\times 40$
  links with oil saturation $S=0.3$ and $\text{Ca}=10^{-1}$ here. The
  initial part of the plot ($N<200$) shows the change in fractional
  flow in an oil-wet system where it approaches to a steady state with
  $F\approx 0.235$. The wettability alteration starts at $N=200$,
  where results for simulations with different values of $\eta$ are
  plotted in different colors. The system evolves to a new
  steady-state with a higher average fractional flow, which is same
  for any value of $\eta$.  $\eta$ only has an effect on the rate of
  change in $F$, which is shown in the inset. A higher value of $\eta$
  results in a longer time to reach the new steady-state.}
\end{figure}

\begin{figure}
\centerline{\hfill
\psfig{file=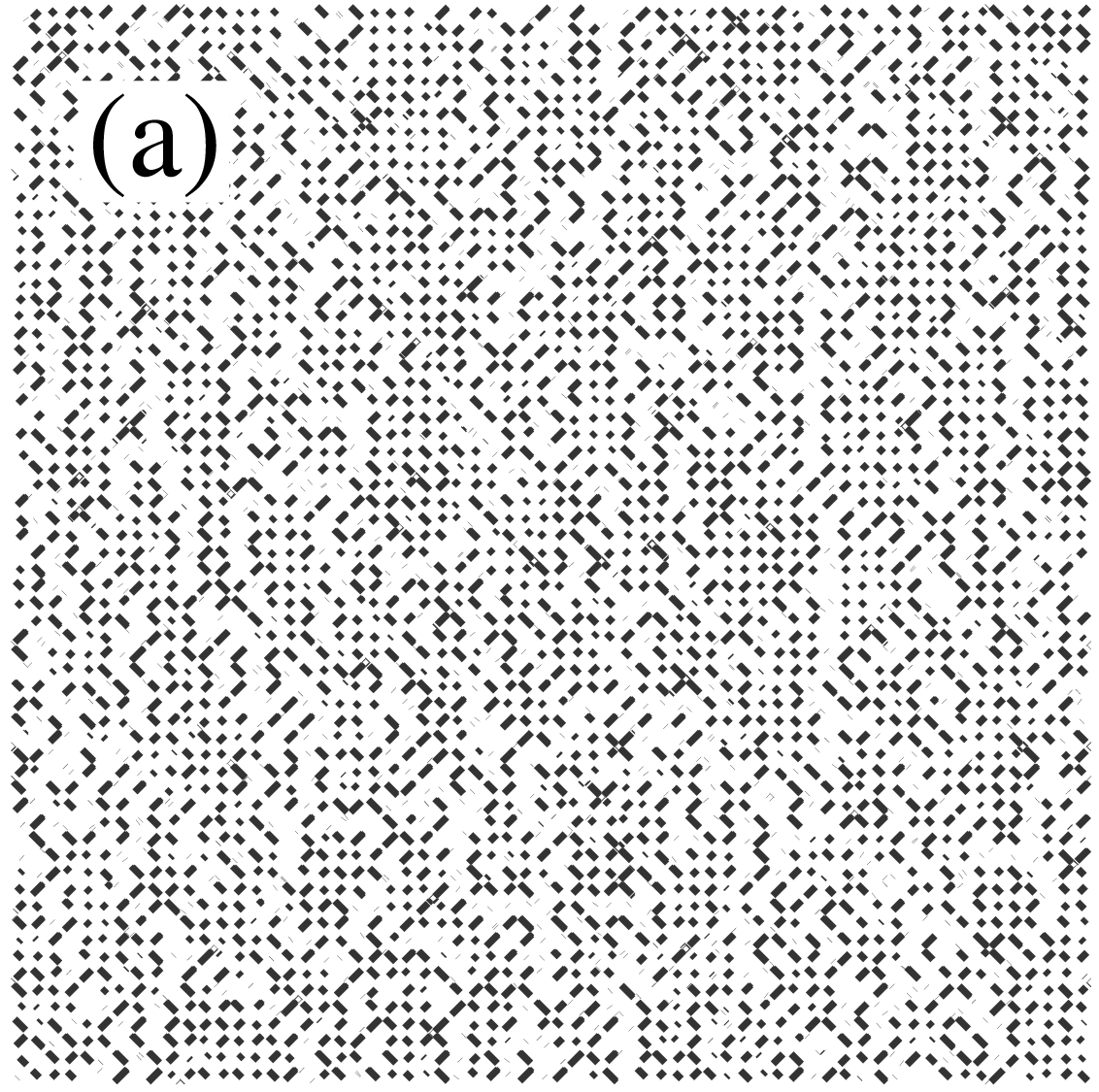,width=0.18\textwidth}\hfill
\psfig{file=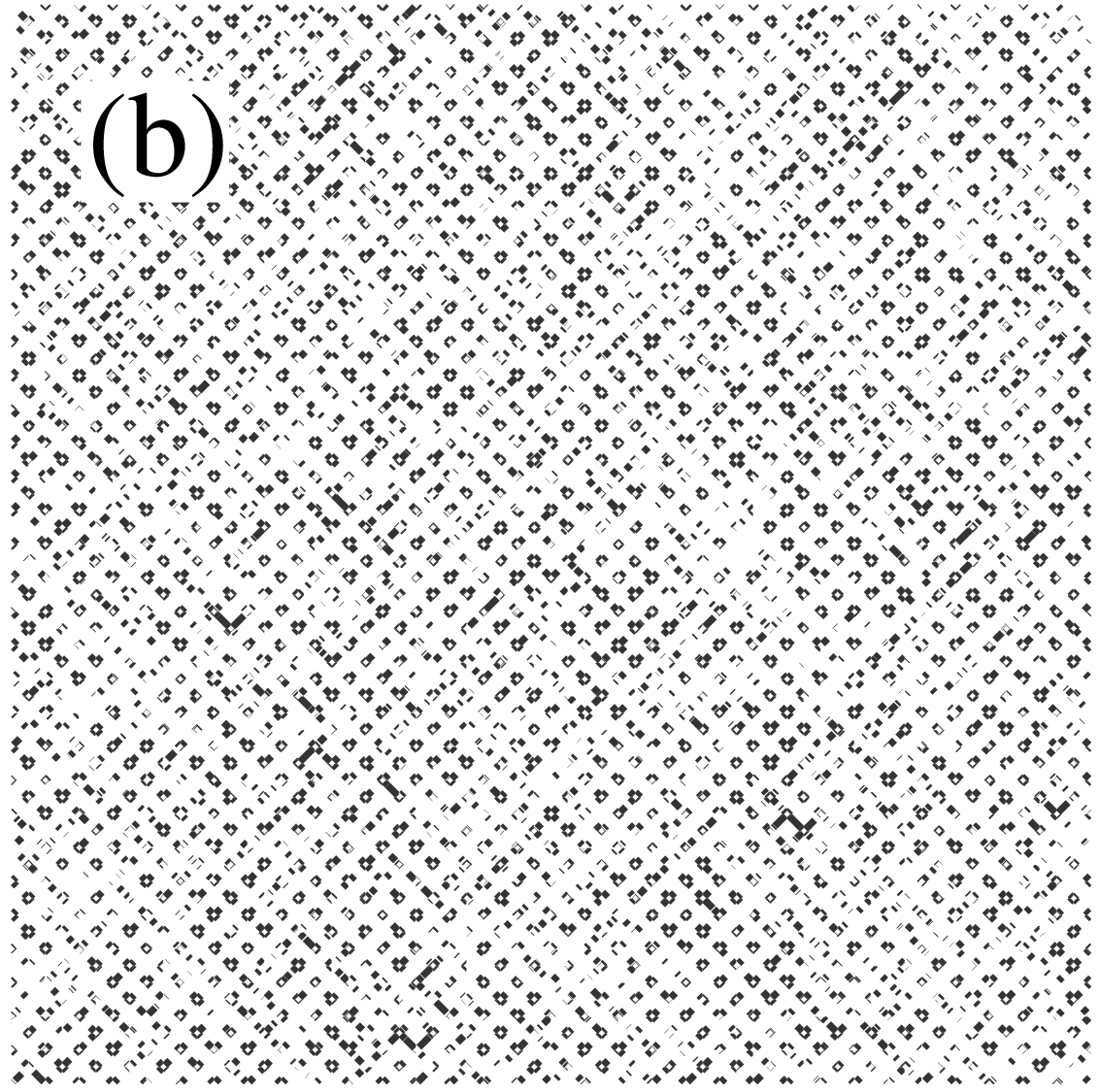,width=0.18\textwidth}\hfill
\hspace{0.05\textwidth\hfill}}
\medskip
\centerline{\hfill
\psfig{file=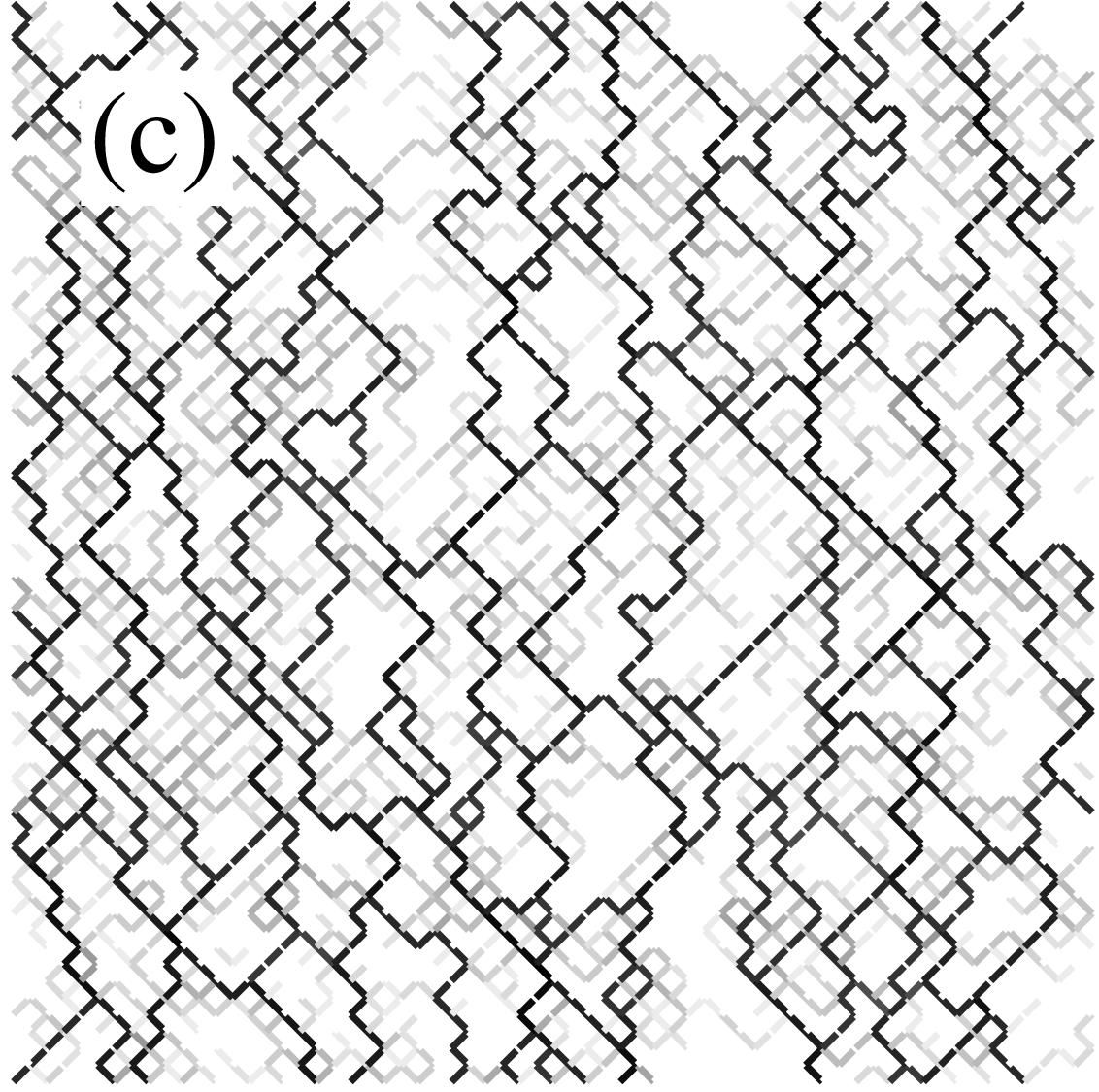,width=0.18\textwidth}\hfill
\psfig{file=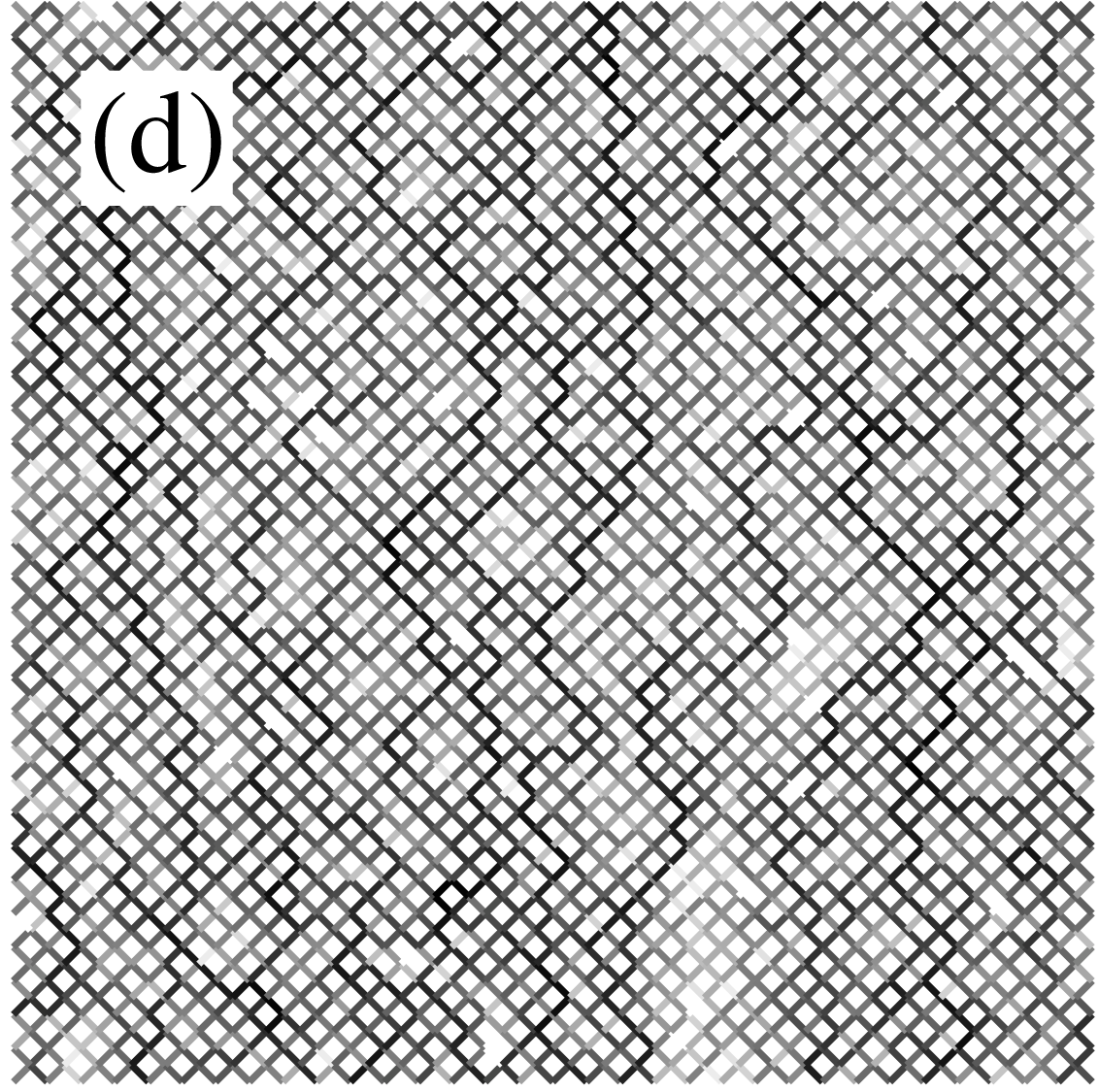,width=0.18\textwidth}\hfill
\psfig{file=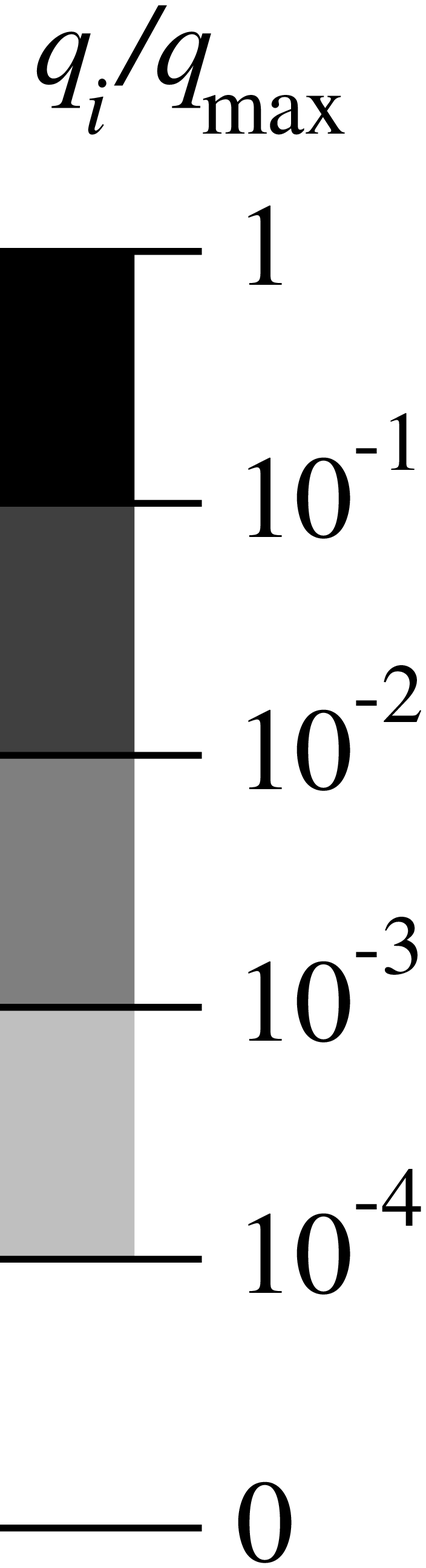,width=0.05\textwidth}\hfill}
\caption{\label{snapshot} Distribution of fluid bubbles (top row) and
  local flow rates (bottom row) in steady state in a network of
  $64\times 64$ links with oil saturation $S=0.3$ and
  $\text{Ca}=10^{-2}$. The left column is for the steady state in a
  oil-wet network and the right column is the new steady state after
  wettability alteration takes place. In (a) and (b), the oil bubbles
  are drawn in black. In (c) and (d), the normalized local flow-rates
  $q_i/q_\text{max}$ are drawn in gray scale.}
\end{figure}

Simulations are started with a random initial distribution of oil and
water in an oil-wet network, where $\theta\approx 165^\circ$ for all
links. First, the oil-wet system is evolved to a steady state before
any wettability alteration is started. This will allow us to compare
the change in the steady-state fractional flow of oil ($F$) with a
change in the wetting angle. The oil fractional flow ($F$) is defined
as the ratio of the oil flow-rate ($Q_\text{oil}$) to the total
flow-rate ($Q$) given by, $F=Q_\text{oil}/Q$. The flow rate ($Q$) is
kept constant throughout the simulation, which sets the capillary
number $\text{Ca} = \mu_\text{eff} Q/(\gamma A)$, where $A$ is the
cross-sectional area of the network. A network of $40\times 40$ links
are considered, which is sufficient to be in the asymptotic limit for
the range of parameters used \cite{meniscimove}. An average over $5$
different realizations of the network has been taken for each
simulation. As the simulation continues, both drainage and imbibition
take place simultaneously due to bi-periodic boundary conditions and
the system eventually evolves to a steady state, with a distribution
of water and oil clusters in the system. In Fig. \ref{eta}, $F$ is
plotted against the number of pore volumes passed ($N$) through the
network. As we run the system with constant flow-rate, $N$ is directly
proportional to the time $t$, $N=tQ/v$ where $v$ is the total volume
of the network. The initial $200$ pore volumes are for an oil-wet
network, where it reaches to a steady state with $F \approx 0.235$. We
then initiate the dynamic wettability alteration which resembles the
flow of a wettability altering brine and $F$ starts to drift. Here we
run simulations for different values of $\eta$, defined in
Eq. \ref{vth}, and the results are plotted in different
colors. $\theta = 0^\circ$ in these simulations, which means any pore
can change to pure water-wet depending upon the flow of brine through
it. One can see that $F$ approaches to a new steady-state with
$F\approx 0.308$ due to the wettability alteration. Here different
values of $\eta$ make the simulation faster or slower to reach the
same steady state, where a higher value of $\eta$ makes the system
slower to reach the steady state. The initialization of steady state
measured in terms of the pore-volumes ($N_\text{ss}$) is defined as
the instant when the average fractional flow stops changing with time
anymore and essentially stays within its fluctuation. In the inset of
Fig. \ref{eta}, $N_\text{v}^\text{ss}$ is plotted with $\eta$ and a
simple linear dependency is observed. In order to save computational
time, we therefore use $\eta=10$ in all our following simulations.

\begin{figure}
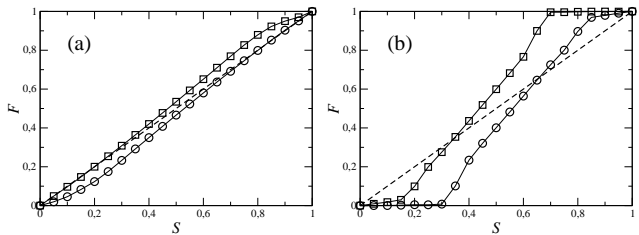

\centerline{\hfill
\psfig{file=fracFlowLvsS_L-0040_M-1.0e+00_Ca-1.0e-01_4m_T-000.0_E-10_V-200-200_iB.eps,width=0.23\textwidth}\hfill
\psfig{file=fracFlowLvsS_L-0040_M-1.0e+00_Ca-1.0e-02_4m_T-000.0_E-10_V-400-400_iB.eps,width=0.23\textwidth}\hfill}
\caption{Oil fractional-flow ($F$) in the steady state at different
  oil saturation ($S$) in the initial oil-wet system ($\bigcirc$) and
  after the wettability alteration with $\theta_\text{c}=0^\circ$
  ($\Box$). Individual simulations are performed for all $S$ values at
  two different capillary numbers (a) $\text{Ca}=10^{-1}$ and (b)
  $\text{Ca}=10^{-2}$. $F$ is higher after wettability alteration for
  the whole range of $S$.}
\label{fracflow0}
\end{figure}

How the two fluids and the local flow-rates are distributed in the
network in the two steady-states before and after the wettability
alteration are shown in Fig. \ref{snapshot}. The network size is
$64\times 64$ links here with an oil-saturation $S=0.3$ and the
capillary number $\text{Ca}=10^{-2}$. All the links are hour-glass
shaped in the actual simulation with disorder in radii, but shown as a
regular network for simplicity in drawing. The upper row shows the
distribution of oil bubbles drawn in black. The left column (a) shows
the steady state in a oil-wet network and the right column (b) shows
the steady-state after the wettability alteration is initiated with
maximum possible wetting angle change $\theta=0^\circ$. A closer look
in these bubble distributions shows more clustered oil bubbles in (a)
than in (b) where they are more fragmented. A more interesting picture
can be seen in the local flow-rate distribution in the bottom row,
which shows a more dynamic scenario. The left (c) and right (d)
figures are for the same time-steps before and after wettability
alteration as in (a) and (b) respectively. Here the local flow-rates
in each pore, normalized in between $0$ to $1$, are shown in gray
scale. Interestingly, in the oil-wet system (c), the flow is dominated
in a few channels (black lines) where the flow-rates are orders of
magnitude higher than the rest of the system. Other than those
channels, the system has negligible flow, indicated by white patches
which means the fluids are effectively stuck in all those areas. This
situation happens when the difference in the saturation of the two
fluids is large, where the phase with higher saturation (water here)
tries to percolate in paths dominated by a single phase with less
number of interfaces. This is not favorable in oil-recovery, as it
leaves immobile fluid in the reservoir. In the flow distribution after
the wettability alteration (d), the flow is more homogeneous and
distributed over the whole system, indicating higher mobility of the
fluids. However, one should remember that when the wettability
alteration is started in a system shown in (c), the wettability
alteration starts taking place only in those pores with active
flow. But then it perturbs the flow-field and starts new flow paths
and eventually the system drifts towards a more homogeneous flow with
time, as shown in (d).

\begin{figure}
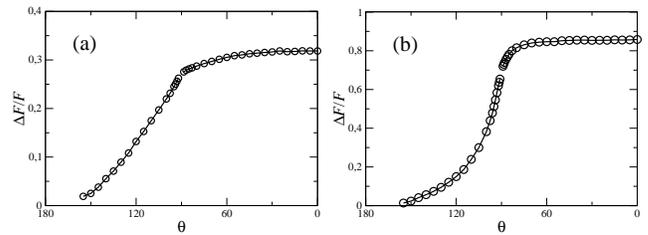

\centerline{\hfill
\psfig{file=fracFlowLvsT_L-0040_M-1.0e+00_Ca-1.0e-01_S-0.30_4m_E-10_iB.eps,width=0.23\textwidth}\hfill
\psfig{file=fracFlowLvsT_L-0040_M-1.0e+00_Ca-1.0e-02_S-0.40_4m_E-10_iB.eps,width=0.23\textwidth}\hfill}
\caption{\label{dF_T0} Proportionate change in the steady-state oil
  fractional-flow ($\Delta F/F$) due to wettability alteration as a
  function of maximum wetting angle $\theta$ for (a)
  $\text{Ca}=10^{-1}$ with $S=0.3$ and (b) $\text{Ca}=10^{-2}$ with
  $S=0.4$.}
\end{figure}

We now present the results when the wetting angle of any pore can
change all the way down to zero degree ($\theta=0^\circ$). In
Fig. \ref{fracflow0} the steady-state oil fractional-flow in an
initial oil-wet system ($F$) is compared with that in the steady-state
after wettability alteration ($F'$). Results are plotted as a function
of $S$ for two different capillary numbers, (a) $\text{Ca}=10^{-1}$
and (b) $\text{Ca}=10^{-2}$. The diagonal line corresponds to $F=S$. A
miscible fluid mixture would follow this line, and it is interesting
to use this as a reference to compare how the data points lie above or
below this line. If both the fluids flow equally in the system then
$F$ will be equal to $S$. But in Fig. \ref{fracflow0}, $F$ stays below
the $F=S$ line for low oil saturation values in the oil-wet system,
which means that the flow of oil is less than that of water and there
are stuck region of oil. A significant increase in $F$ can be observed
here due to the wettability alteration for the full range of
oil-saturation. Moreover, increase in $F$ is higher for the lower
capillary number, indicating that wettability alteration is very
significant in the case of oil recovery, as $\text{Ca}$ can go as low
as $10^{-6}$ in the reservoir pores. Fractional flow also obeys the
symmetry relation $F'(S)=1-F(1-S)$ \cite{sinha11} which implies that,
if the wetting angle of any pore is allowed to change all way down to
zero degree ($\theta=0^\circ$), the system will eventually become pure
water-wet with time.

\begin{figure}
\centerline{\hfill
  \psfig{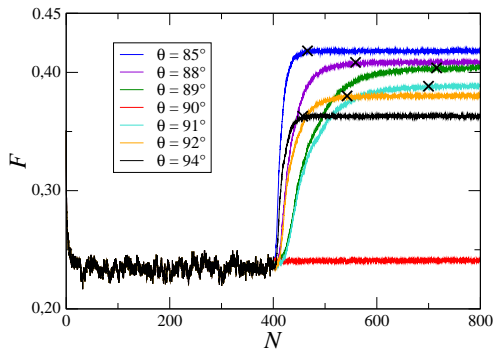}\hfill}
\caption{\label{FvN-T}Change in $F$ during the simulation for
  different $\theta$ values. The wettability alteration started at
  $N=400$ pore-volumes. The initialization of the steady-states for
  different $\theta$ values are marked by crosses on the respective
  plots.}
\end{figure}

As noted earlier, the maximum change in wetting angle for a system,
depends on the properties of the reservoir rock, crude oil and brine,
and also on the temperature. Existing wettability alteration
procedures generally turns the oil-wet system into intermediate wet,
rather than to pure water-wet. Some examples of the change in the
wetting angle for different rock materials and brine can be found in
\cite{wang01, wang02}. In our simulation this is taken care of by the
parameter $\theta$, which decides the maximum change in the wetting
angle $\vartheta_i$ for any pore. One should remember that, we are not
forcibly changing the wetting angles $\vartheta_i$ to $\theta$, rather
the change in $\vartheta_i$ is decided independently for individual
pores by the amount of brine passed through it (by Eqs. \ref{flowsum}
to \ref{pcmod}), and there is a maximum allowed change in any
$\vartheta_i$.  As before, simulations are started with a pure oil-wet
system to reach a steady state and then wettability alteration is
started and simulation continues until the system reaches to a steady
state again. Independent simulations have been performed for different
values of $\theta$. The proportionate change in the oil
fractional-flow due this wettability change from the oil-wet system,
$\Delta F/F=(F'-F)/F$ is measured for different simulations and
plotted in Fig. \ref{dF_T0} against $\theta$. There are a few things
to notice. First, as one can immediately see, fractional flow
increases with the decrease of oil-wetness, $\theta\to 0^\circ$. The
maximum increase in $F$ is higher for lower $\text{Ca}$, about $86\%$
for $\text{Ca}=10^{-2}$ and about $32\%$ for $\text{Ca}=10^{-1}$. This
is because the change in wetting angle affects the capillary pressures
at the menisci, so the change in $F$ is larger when the capillary
forces are higher. Secondly, the major change in $F$ happens in the
intermediate wetting regime, upto $\theta\approx 60^\circ$, and then
it becomes almost flat afterwards. Moreover, this increase in $F$ is
more rapid for lower value of $\text{Ca}$. All these facts points
towards an optimal range of wetting angle change to increase the oil
flow. This is an important observation for practical reasons, as it is
not necessary to change the wetting angle further. Thirdly, there is a
discontinuity in the curve exactly at $\theta=90^\circ$, as we will
discuss later.

\begin{figure}
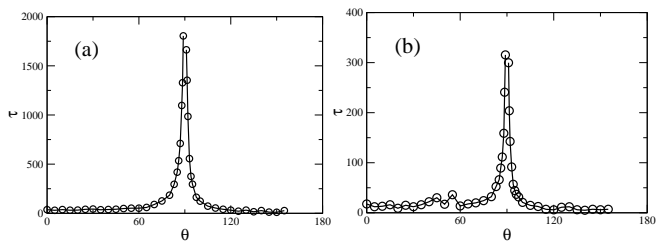

\centerline{\hfill
\psfig{file=timeStedyvsT_L-0040_M-1.0e+00_Ca-1.0e-01_S-0.30_4m_E-10_iB.eps,width=0.24\textwidth}\hfill
\psfig{file=timeStedyvsT_L-0040_M-1.0e+00_Ca-1.0e-02_S-0.30_4m_E-10_iB.eps,width=0.24\textwidth}\hfill}
\caption{\label{NvT}Variation of steady-state initialization time
  ($\tau$) as a function of $\theta$ for (a) $\text{Ca}=10^{-1}$ with
  $S=0.3$ and (b) $\text{Ca}=10^{-2}$ with $S=0.4$. $\tau$ diverges
  rapidly as $\theta\to\theta_c$ where $\theta_c=90^\circ$.}
\end{figure}

Increase in the oil fractional-flow with the increase in the
water-wetness may seem to be obvious and reported by many experiments
and field tests. But, the most important concern for the oil industry
is the rate of increase, or the time required to achieve a significant
increase in the oil production. If the increment in oil flow is very
slow compared to the cost of the process, then the oil recovery is
declared as not profitable and the reservoir may be considered to be
abandoned. As per our knowledge, there are very few systematic studies
reported in the literature predicting the time scale to change the oil
flow due to the wettability change by two-phase flow of brine and oil
in a porous media. We observe that, due to the correlations between
the flow paths and the wetting angle change, the time scale of the
process varies dramatically with $\theta$. This is illustrated in
Fig. \ref{FvN-T}, where $F$ is plotted as a function of pore volumes
($N$) of fluid passed through the system. The initial $400$ pore
volumes are for an oil-wet system and then results of few different
simulations with $\theta=85$, $88$, $89$, $90$, $91$, $92$ and $94$
degrees are plotted. Interestingly, the rate at which the system
reaches a new steady-state, varies significantly depending on the
value of $\theta$. For example, after the wettability alteration is
started, it needs to flow less than $100$ pore volumes to reach the
new steady state for $\theta=94^\circ$ whereas more than $300$ pore
volumes are needed to reach a steady state for
$\theta=91^\circ$. Therefore, even if the final steady-state
fractional flow is higher for $\theta=91^\circ$ than for
$\theta=94^\circ$, it might not be profitable to alter the wetting
angles to $91^\circ$ because of the slow increase in $F$. In general,
the process becomes slower and slower as $\theta\to90^\circ$ from both
sides. Such kind of slow increase in oil recovery as
$\theta\to90^\circ$ is also observed in experiments
\cite{wang01,wang02}. This slowing down of the process is an combined
effect of two factors. First, the fact that wettability only can
change in the pores where there is flow of brine and the second is the
value of $\theta$. All the pores were initially oil-wet
($\vartheta_i\approx165^\circ$) and when it reaches the steady state,
the flow finds the high mobility pathways depending on the mobility
factor of the pores and the capillary pressures at the menisci. When
the wettability alteration is started, the wetting angles of the
existing flow pathways start decreasing. As a result, capillary
pressures at menisci in those channels first decreases as
$\vartheta_i\to90^\circ$ and then it increase afterwards as
$\vartheta_i\to0^\circ$. This creates a perturbation in the global
pressure field and correspondingly viscous pressures start changing
with time which changes the flow field. However, capillary pressures
at the zero-flow regimes are now higher than the high-flow regimes
which makes it difficult to invade the zero-flow regimes causing a
slower change in the flow field as $\vartheta_i$ approaches
$90^\circ$. An interesting feature is observed exactly at
$\theta=90^\circ$, where the average fractional flow does not change
at all after the wettability change. At exactly $\theta=90^\circ$,
capillary pressures in all the pores in the existing flow pathways
essentially become zero, making them the lowest resistive channels
with zero capillary barriers. As a result, the fluids keep flowing in
the existing channels forever and the system stays in the same
steady-state. The time taken to reach another new steady-state is
therefore infinite at $\theta=90^\circ$ and it therefore is a critical
point for the system.

\begin{figure}
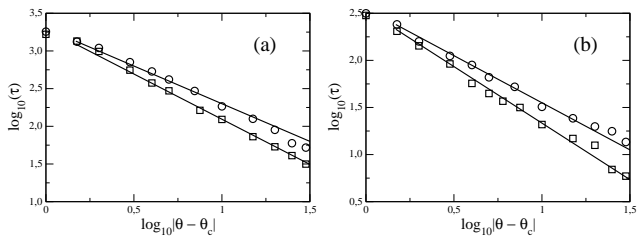

\centerline{\hfill
\psfig{file=timeStedyvsT_L-0040_M-1.0e+00_Ca-1.0e-01_S-0.30_4m_E-10_iB-log.eps,width=0.23\textwidth}\hfill
\psfig{file=timeStedyvsT_L-0040_M-1.0e+00_Ca-1.0e-02_S-0.30_4m_E-10_iB-log.eps,width=0.23\textwidth}\hfill}
\caption{\label{NvT-Log}Plot of $\log\tau$ vs
  $\log\lvert\theta-\theta_c\rvert$ for (a) $\text{Ca}=10^{-1}$ with
  $S=0.3$ and (b) $\text{Ca}=10^{-2}$ with $S=0.4$. From the slopes,
  the value of the dynamic critical exponent $\alpha$ is obtained as
  $\alpha=1.0\pm 0.1$ for $\theta<\theta_c$ ($\bigcirc$) and
  $\alpha=1.2\pm 0.1$ for $\theta>\theta_c$ ($\Box$).}
\end{figure}

We now measure the steady-state initialization time $\tau$, defined as
the moment when the average of the fractional flow stops changing with
time and becomes horizontal with the $x$ axis. This is shown In
Fig. \ref{FvN-T}, where the initialization of steady states is marked
by crosses on the respective plots. As the simulations are performed
with constant $Q$, $\tau$ is proportional to the fluid volume passed
through the system and therefore we measure $\tau$ in the units of
$N$. $\tau$ for different simulations with different values of the
maximum wetting angle ($\theta$) is plotted in Fig. \ref{NvT} for (a)
$\text{Ca}=10^{-1}$ with $S=0.3$ and (b) $\text{Ca}=10^{-2}$ with
$S=0.4$. One can see that $\tau$ diverges rapidly as $\theta$
approaches $\theta_\text{c}=90^\circ$ from both sides,
$\theta>90^\circ$ and $<90^\circ$. This divergence of the steady-state
time $\tau$ as $\theta\to\theta_\text{c}$ indicates the {\em critical
  slowing down} of the dynamics, which is a characteristics of
critical phenomena. The critical slowing down is the outcome of the
divergence of correlations at the critical point and can be
characterized by a dynamic critical exponent $z$ defined as
$\tau\sim\xi^z$, where $\xi$ is the correlation length
\cite{noneqthr}. As $\theta\to\theta_\text{c}$, the correlation length
$\xi$ diverges as $\lvert\theta-\theta_\text{c}\rvert^{-\nu}$ and
therefore $\tau\sim\lvert\theta-\theta_{c}\rvert^{-\alpha}$, where
$\alpha=z\nu$. In Fig. \ref{NvT-Log}, $\tau$ is plotted versus
$\lvert\theta-\theta_\text{c}\rvert$ in log-log scale which gives two
different slopes for $\theta>\theta_\text{c}$ and
$\theta<\theta_\text{c}$. We then find the value of the dynamic
exponents $\alpha$ as $\alpha=1.2\pm0.1$ for $\theta>\theta_\text{c}$
and $\alpha=1.0\pm0.1$ for $\theta<\theta_\text{c}$. However, they are
the same within error bar for different capillary numbers and
saturations (Fig. \ref{NvT-Log}). The value of the dynamic critical
exponents depend on the underlying dynamics and on the model
\cite{crslowising}. In this case, wettability alteration was started
from an oil-wet system with $\vartheta_i=165^\circ$ for all the
pores. So for the simulations with $\theta<90^\circ$, the wetting
angles cross the critical point ($90^\circ$) when the capillary forces
change directions. This might cause the system to mobilize the
clusters somewhat faster than for $\theta>90^\circ$ when the capillary
forces does not change any direction. As a result, $\alpha$ becomes
smaller for $\theta<90^\circ$ than for $\theta>90^\circ$.

\section{Conclusions}\label{Conclusions}
In this article we have presented a detailed computational study of
wettability alterations in two-phase flow in porous media, where the
change in the wetting angle in a pore is controlled by the volumetric
flow of the altering agent through it. When the wetting angles are
allowed to alter towards water-wetness, the stuck oil clusters start
to mobilize and oil-fractional flow increases. However, due to the
correlations in the wetting angle change with the flow pathways, the
time-scale of the dynamics strongly depends on the maximum allowed
change in the wetting angle. We find that, as the final wetting angle
is chosen closer to $90^\circ$, the system shows a critical slowing
down in the dynamics. This critical slowing down is characterized by
two dynamic critical exponents.  The critical point we are dealing
with is an equilibrium critical point as the system is in steady
state.  The dynamical critical exponents measure how long it takes to
go from one steady state to a new one. To our knowledge, this is the
first example of there being {\it different values\/} for the
exponents on either side of the critical point.  Our findings are in
agreement with experimental observations reported in literature, and
are extremely important for application purposes like oil recovery,
where the time scale of the process is a key issue.

\section{Acknowledgements}
We thank E.\ Skjetne for introducing us to the subject of this study.
We have benefited from discussions with D.\ Bideaux,
E.\ G.\ flekk{\o}y, S.\ Kjelstrup and K.\ J.\ M{\aa}l{\o}y. This work
has been supported by the Norwegian Research Council.  We furthermore
thank the Beijing Computational Sciences Research Center and its
Director, H.\ Q.\ Lin for hospitality during the final stages of this
work.

\end{document}